\documentclass[12pt]{article}
\pdfoutput=1

\usepackage[utf8]{inputenc}
\usepackage[tbtags]{amsmath} 				
\usepackage{amssymb, mathrsfs}			
\usepackage{physics}
\usepackage{graphicx}			
\usepackage{tensor}				
\usepackage[makeroom]{cancel}	
\usepackage{bm}								
\usepackage[
      colorlinks=true,
      linkcolor=blue,
      urlcolor=blue,
      filecolor=black,
      citecolor=red,
      pdfstartview=FitV,
      pdftitle={},
      pdfauthor={Michael Gutperle, Nicholas Klein},
      bookmarksopen=true
      ]{hyperref}
\usepackage{calc}
\usepackage{sectsty}
\usepackage{dsfont}
\usepackage{cite}

\newlength\myheight
\newlength\mydepth
\settototalheight\myheight{Xygp}
\sectionfont{\large}

\renewcommand{\thefootnote}{\fnsymbol{footnote}}
\renewcommand{\thanks}[1]{\footnote{#1}}
\newcommand{\starttext}{
\setcounter{footnote}{0}
\renewcommand{\thefootnote}{\arabic{footnote}}}
\renewcommand{\epsilon}{\varepsilon}	

\numberwithin{equation}{section} 		

\numberwithin{equation}{section}

\long\def\symbolfootnote[#1]#2{\begingroup%
\def\thefootnote{\fnsymbol{footnote}}\footnote[#1]{#2}\endgroup}

\begin{document}
\setlength{\baselineskip}{16pt}

\starttext
\setcounter{footnote}{0}

\begin{flushright}
\today
\end{flushright}

\bigskip

\begin{center}

{\Large \bf  A note  on co-dimension 2 defects in $N=4,d=7$  gauged supergravity}

\vskip 0.4in

{\large   Michael Gutperle  and  Nicholas Klein }

\vskip 0.2in

{\sl Mani L.~Bhaumik Institute for Theoretical Physics}\\
{\sl Department of Physics and Astronomy }\\
{\sl University of California, Los Angeles, CA 90095, USA}

\end{center}
 
\bigskip
 
\begin{abstract}
\setlength{\baselineskip}{16pt}

In this note we present a solution of $N=4,d=7$ gauged supergravity which is holographically dual to a co-dimension two defect living in a six dimensional SCFT. The solution is obtained by double analytic continuation of a two  charge 
supersymmetric  black hole solution. The condition that no conical deficits are present in the bulk and on the boundary  is satisfied by a one parameter family of solutions for which some holographic observables are computed.

\end{abstract}

\setcounter{equation}{0}
\setcounter{footnote}{0}

\newpage

\section{Introduction}

The construction and study of extended conformal defects  is an important subject  in the investigation of superconformal field theories (SCFT).    Defects are characterized by the  broken and preserved symmetries. In  a $d$-dimensional  SCFT, a $p$-dimensional conformal defect  preserves a $SO(p,2)\times SO(d-p)$ subgroup of the $SO(d,2)$ conformal group. The first factor is the conformal symmetry acting on the world volume of the defect and the second factor is the rotational symmetry in the transverse directions, which acts like a global symmetry on the degrees of freedom localized on the defect.

If the SCFT has a holographic dual it is interesting to look for the holographic description of such defects, which fall into two categories: First, a brane is placed in the bulk spacetime which ends on the boundary at the $p$ dimensional defect \cite{Karch:2000gx,DeWolfe:2001pq}.  In a probe approximation the gravitational back reaction of such the brane is neglected, but the embedding is determined by solving the world volume equations of motion or the BPS-condition following from world volume kappa symmetry \cite{Simon:2011rw}.  Second, a fully back reacted solution of the supergravity can be constructed using an ansatz of $AdS$ and sphere factors warped over a base space (which can be a line or a Riemann surface with boundary). Solutions can either be constructed in lower dimensional gauged supergravities \cite{Clark:2005te,Bobev:2013yra} and in favorable circumstances be uplifted  ten or eleven dimensions,  or alternatively solutions  can be constructed in ten or eleven dimensions \cite{Bak:2003jk,DHoker:2007zhm,DHoker:2007mci,DHoker:2008lup}.  The former solutions are easier to obtain but the later are more general and in many cases give a top down  understanding of the defects as backreacted solutions of intersecting brane systems, which allow us to identify the gauge theories, often of quiver type,  which flow to the SCFTs.

In this note we consider the holographic description of $p=4$ dimensional defects in $d=6$ dimensional SCFTs. We construct  solutions in a  truncation of maximal $SO(5)$ gauged supergravity in seven dimensions with $U(1)\times U(1)$ gauge symmetry.  These solutions are related by a double  analytic continuation to supersymmetric black hole solutions.  They are also closely related to compactifications of the seven dimensional theory on spindles - two dimensional compact surfaces with  conical deficits which have been studied extensively in the past two years (see e.g.\cite{Ferrero:2020laf,Ferrero:2021wvk,Ferrero:2021etw,Faedo:2021nub,Couzens:2021rlk,Hosseini:2021fge}).   Both  constructions  start with a ansatz $AdS_5\times S^1$ warped over a real coordinate.  Whereas the spindle solution the real coordinate takes values on a compact interval and the circle closes off at either end of the interval, in our case the real coordinate takes values on a real half-line and the geometry decompactifies to an asymptotic $AdS_7$ space. The solution therefore describes conformal a defect living inside  a higher dimensional SCFT.

The structure of this note  is as follows. In section \ref{sec:two} we describe the seven dimensional gauged supergravity and the relevant solutions  which are obtained from double analytic continuation of black hole solutions. In section \ref{sec:three} we perform a regularity analysis based on the absence of conical singularities in the bulk and boundary and obtain a one parameter family of regular solutions, as well as solutions with conical singularities in the bulk related to spindles which have been actively investigated recently.
In section \ref{sec:four} we perform some holographic calculations using the regular solutions, in particular we calculate the on-shell action of the solution, as well as the expectation value of the stress tensor and conserved R-symmetry currents.  In section \ref{sec:five} we briefly discuss the uplift of the solution to eleven dimensions which is used to identify the R-symmetry currents  of the six dimensional SCFT  to which the seven dimensional  gauge fields  are dual. We close with a discussion of our results and leave some details of calculations to an appendix.

\section{7-dim gauged supergravity}
\label{sec:two}

We consider a  truncation of maximal $N=4$, $SO(5)$ gauged supergravity in seven dimensions  \cite{Pernici:1984xx} with $U(1)\times U(1)$ gauge symmetry and two scalars \cite{Cvetic:1999xp,Lu:2003iv,Liu:1999ai}.  There exists a consistent uplift of the seven dimensional solutions to eleven dimensional supergravity \cite{Cvetic:1999xp}.  The solutions we consider are double analytic solutions of 
charged non-rotating black hole solutions \cite{Liu:1999ai,Lu:2003iv}, where the $S^5$ factor  is replaced by a $AdS_5$ factor and the time coordinate is replaced by a space-like compact circle coordinate.  The black hole solution depends on a non-extremality parameter and two charges. The extremal solution preserves  either half or a quarter of the thirty-two supersymmetries of the gauged supergravity theory for one or two nonzero charges respectively \cite{Liu:1999ai}.
It was shown in \cite{Ferrero:2021wvk} that the analytically continued extremal solutions also preserve the same amount of supersymmetry.

We follow the conventions of \cite{Ferrero:2021wvk} to facilitate a comparison with their analysis.  The  action for the bosonic fields of $U(1)\times U(1)$ gauged supergravity in seven dimensions  is given by
\begin{align}\label{sugract}
S&= -{1\over 16 \pi G_N} \int d^7 x \sqrt{-g} \Big( R- g_c^2 V(\phi)- {1\over 2} \sum_{i=1}^2 \partial_\mu \phi_i \partial^\mu \phi_i - {1\over 4}  e^{\sqrt{2} \phi_1 + \sqrt{2\over 5} \phi_2 }F_1^2\nonumber\\
&  \hskip1.7in   -{1\over 4}  e^{-\sqrt{2} \phi_1 + \sqrt{2\over 5} }\phi_2 F_2^2 \Big)
\end{align}
where $F_i=dA_i, i=1,2$ and the potential for the scalar fields  is given by
\begin{align}\label{scalpot}
V(\phi)={2g_c^2} e^{-\sqrt{2\over 5} \phi_2}\Big( -8 + e^{\sqrt{10}\phi_2}-8 e^{\sqrt{5\over 2} \phi_2 }\cosh{\phi_1\over \sqrt{2}}\Big)
\end{align}
The solution given in \cite{Ferrero:2021wvk} can be expressed in term of the following functions
\begin{align}\label{fundef}
h_i(y) &= y^2+q_i, \quad i=1,2\nonumber \\
P(y)&= h_1(y) h_2(y) \nonumber \\
Q(y)&= -y^3-\mu y+  g_c^2 h_1(y) h_2(y)
\end{align}
and is given by
\begin{align}\label{2dmetric}
ds^2&=\Big( y P(y)\Big)^{1\over 5} \Big\{ ds_{AdS_5}^2+ {y\over 4 Q(y)} dy^2 +{Q(y)\over P(y)} dz^2\Big\}\nonumber \\
A_i &=\left( {\sqrt{1-{\mu\over q_i}} q_i\over h_i(y)}\right)dz, \quad i=1,2\nonumber \\
e^{\phi_1}& = \left( {h_1(y)\over h_2(y)}\right)^{1\over \sqrt{2}}, \quad e^{\phi_2}= { (h_1(y)h_2(y) )^{1\over \sqrt{10}} \over y^{2\sqrt{2\over 5}}}
\end{align}
 It is easy to verify that the equations of motion following from the variation of the action (\ref{sugract}) are satisfied for such a solution.
Here $q_1,q_2$ are related to the charges and  $\mu$ is a non-extremality  parameter which we set to $\mu=0$. This choice corresponds to a supersymmetric solution. We will also set $g_c=1$ for simplicity. For these choices the solution with $q_1=q_2=0$ corresponds to a unit radius $AdS_7$,  using  $AdS_5\times S^1$ slicing coordinates.  
\section{Regularity analysis}
\label{sec:three}

In this section we present the conditions that regularity imposes on the solution. The analysis follows the general strategy employed in other cases of holographic description of defects. \cite{Chen:2019qib,Gutperle:2019dqf,Chen:2020mtv}. It is also closely related to the construction of holographic calculations of Renyi-entropies \cite{Crossley:2014oea,Hosseini:2019and}, compactifications on spindles  \cite{Ferrero:2021wvk,Ferrero:2020laf,Ferrero:2021etw,Faedo:2021nub} and related constructions \cite{Bah:2021mzw,Bah:2021hei}.

In \cite{Ferrero:2021wvk} the solution presented in section  \ref{sec:two} was used to construct a $AdS_5$ compactification of seven dimensional supergravity  on a two dimensional compact space, a so-called  spindle.   A spindle is   topologically a  two sphere with two conical deficits at the north and south poles respectively.   A spindle exists if the function $Q(y)$, defined in (\ref{fundef}) has two  real zeros and in between the zeros both $Q(y)$ and $P(y)$ are positive. The regularity,   supersymmetry and the quantization of the deficit angle coming from a consistent interpretation of the uplift to eleven dimensions impose conditions on the parameters of the solution  which were worked out in \cite{Ferrero:2021wvk}.

In our case the two dimensional space will be non-compact and  we will look at the region from the largest  positive zero of $Q(y)$ to infinity, which is a region where Q is positive. In the following we will investigate the regularity conditions imposed on the solution.
 For convenience we write out the functions which determined the regularity  (recall we have set $\mu=0$).
\begin{align}
    Q(y) &= -y^3 +  (y^2+q_1)(y^2+q_2) = y^4-y^3 + (q_1+q_2)y^2 +{q_1q_2 }\nonumber \\
    P(y) &= (y^2+q_1)(y^2+q_2)
\end{align}

As $y\to \infty$ we approach an asymptotic $AdS_7$ region, with a six dimensional boundary. 
In this limit the metric takes the form
\begin{align}
\lim_{y\to \infty} ds^2 &= y ds_{AdS_5}^2 + y dz^2 + {1\over 4 y^2} dy^2 +\cdots \nonumber \\
&= {d\rho^2\over 4\rho^2}  + {1\over \rho} \Big( ds_{AdS_5}^2 +  dz^2 \Big) + \cdots
\end{align}
where we defined the Fefferman-Graham coordinate $\zeta$ as $y=1/\rho$ and the dots denote sub-leading terms in $y$ and $\rho$,  which are determined in appendix \ref{appa}.  The metric is asymptotic to $AdS_7$,  Since the $z$ direction parameterizes a circle, the holographic boundary of the asymptotic AdS space is of the form $AdS_5\times S^1$. The six dimensional metric on the boundary is given by
\begin{align}
ds_6^2&= {dr ^2 -dt^2-\sum_{i=1}^3 dx_i^2\over r^2} + dz^2 \nonumber\\
&= {1\over r^2} \Big(  d\zeta ^2 -dt^2-\sum_{i=1}^3 dx_I^2  + r^2  dz^2\Big)
\end{align}
which is conformal to $R^{1,5}$ if the coordinate $z$ has periodicity $2\pi$. For a different periodicity of $z$ the  boundary has  a conical singularity at $r=0$. In the standard formulation of AdS/CFT the boundary theory does not have dynamical gravity and hence a co-dimension two defect does not induce a conical deficit,  as a cosmic string would in a gravitational theory.  Consequently the condition of the absence of a conical deficit on the boundary fixes the periodicity of the $S^1$ coordinate $z$ to be $2\pi$.

We now seek conditions on $q_1,q_2$ such that there is at least one positive zero and that it is not a double zero. Once we have such a $y_+$, we can guarantee that in the range $[y_+,\infty ) $ both metric functions $Q(y)>0$ and $P(y) > y^3 > 0$ are positive and the metric is regular.  An important quantity for the nature of the zeros of $Q$ is the discriminant
\begin{align}\label{discriminant}
D= q_1 q_2 \Big( 16(q_1^4+q_2^4)-4(q_1^3+q_2^3)-64 (q_1^3 q_2+ q_1 q_2^3) +96 q_1^2 q_2^2 +132(q_1^2 q_2+ q_1 q_2^2) -27 q_1 q_2\Big)
\end{align}

\begin{figure}  \centering
  \includegraphics[width=90mm]{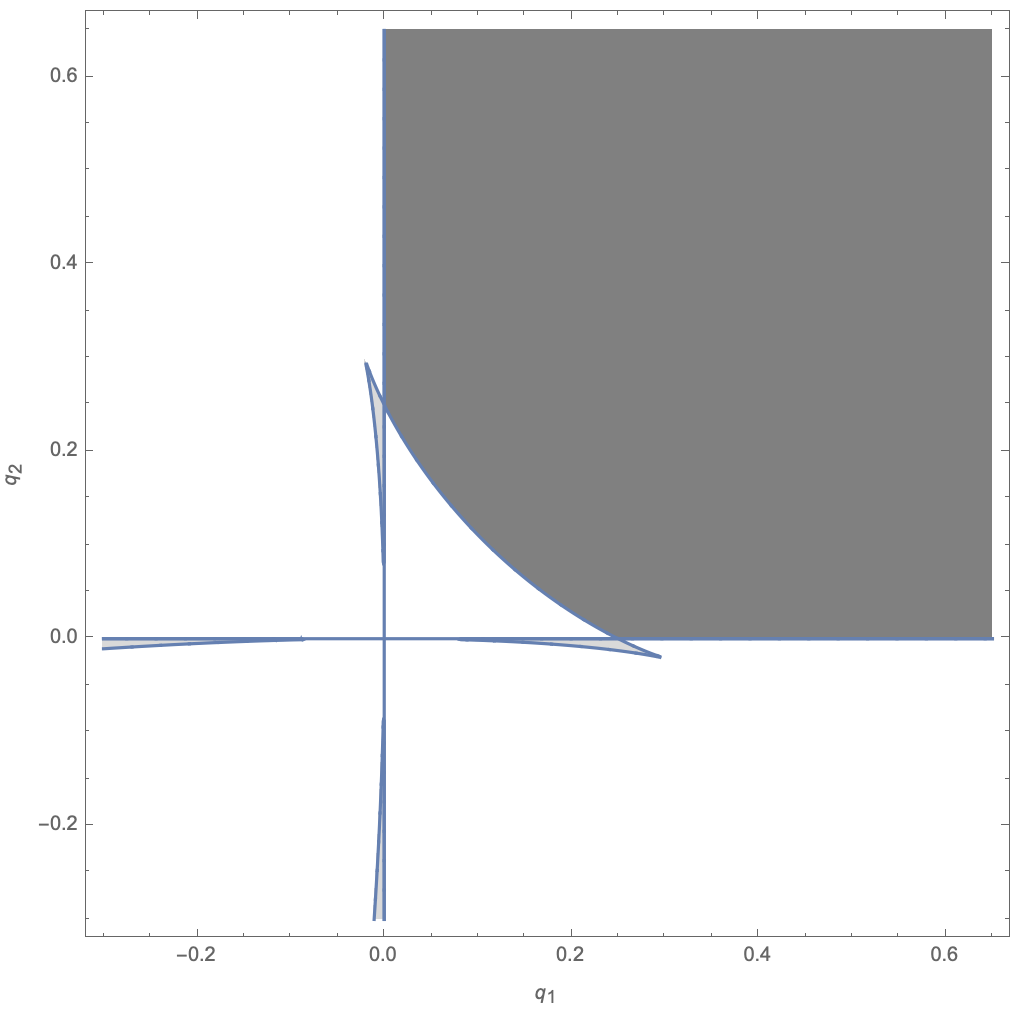}
  \caption{Sign of the discriminant (\ref{discriminant}) of the polynomial $Q(y)$ in the $(q_1,q_2)$ plane }\label{fig:one}
\end{figure}

Note that the vanishing of the discriminant implies the presence of a real double zero and for $D>0$ we have either  four or no real zeros whereas for  $D<0$ we have two real and two complex conjugate roots.  We show a plot of the sign of the discriminant as a function of $q_1,q_2$ in figure  \ref{fig:one}, where locus of vanishing discriminant is represented by the blue curve and regions of positive discriminant are shaded grey.

We can use Descartes' rule of signs to show  that in the region with either one or both $q_1$ and $q_2$ negative,  we have two real roots in the (white) region where $D<0$ and four real roots in the (grey) region where $D>0$. In the region where both $q_1,q_2$ are positive we have two real zeros in the white region where $D<0$ and no real zeros in the (dark grey) region, where $D>0$.   This implies that the dark grey region of charges is excluded since $Q(y)$ is never zero here and we will produce a naked singularity when $y$ goes to zero and the Ricci scalar diverges.

Note that if $y=y_0$ is a double zero the metric will approach the following form near $y= y_0+\rho$
 \begin{align}
 ds^2 \sim (y_0 P(y_0) ^{1\over 5} \Big( ds_{AdS_5}^2+ {y_0 \over   \gamma \rho^2} d\rho^2 + {\gamma  \rho^2 \over P(y_0)} dz^2\Big)
 \end{align}
 where $\gamma ={1\over 2} Q''(y)\mid_{y=y_0}$. This produces a singularity at $\rho=0$. (We will see that we will never have to worry about this case for $q_1,q_2$ which satisfy the other regularity conditions)

\begin{figure}  \centering
  \includegraphics[width=90mm]{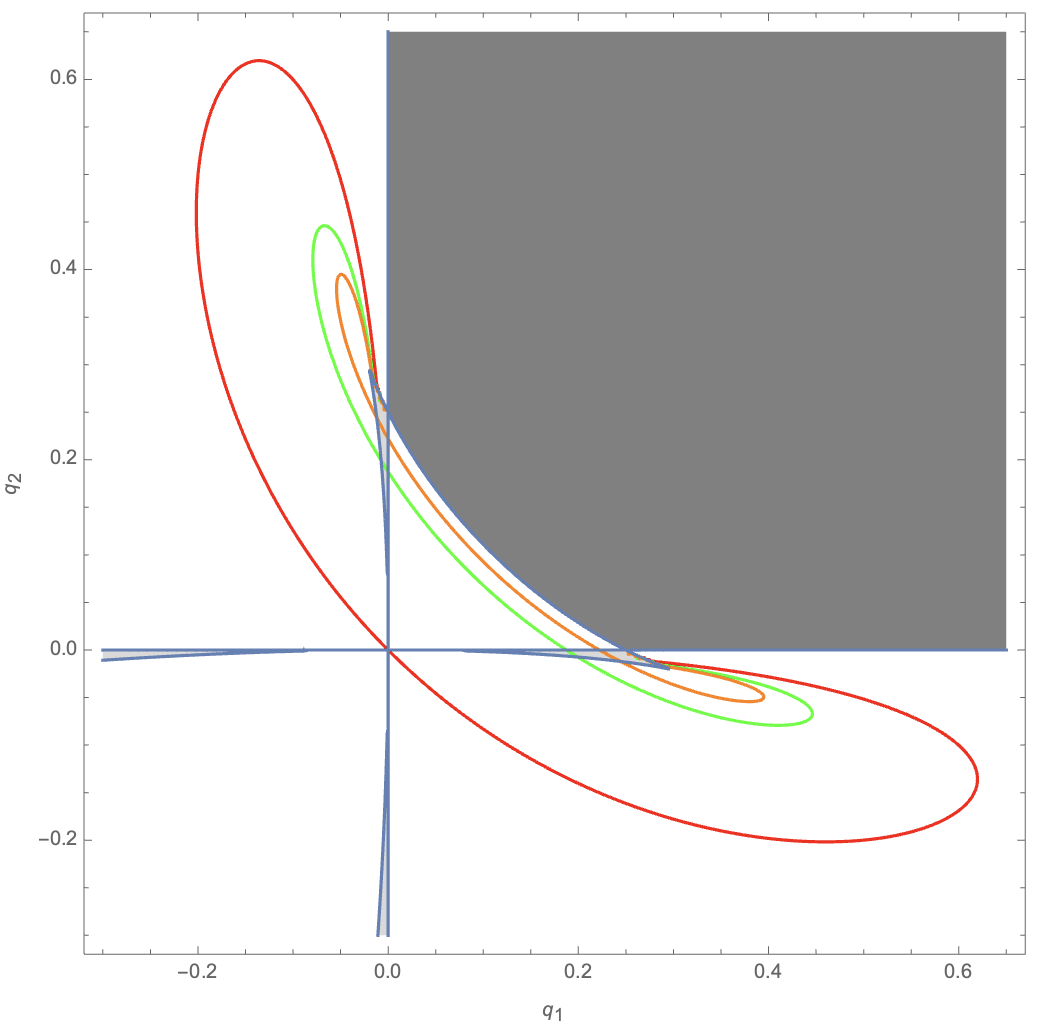}
  \caption{Allowed charges for different values of conical deficits: $n=1$ (red) is the completely regular solution and  two half-spindles with $n=2$ (green) and $n=3$ (orange) }\label{fig:two}
\end{figure}

Now we assume that we are in the allowed region of the $q_1,q_2$ plane and consider the $y\to  y_+$ limit where $y_+$ is the largest positive zero of the function $Q(y)$. Letting $y = y_+ + \rho$, we have that 
\begin{align}
    Q(y) &\approx Q'(y_+)\rho\nonumber\\
    P(y) &\approx P(y_+) = (Q(y_+) + y_+^3) = y_+^3
\end{align}
Plugging these into the metric (\ref{2dmetric}) and defining the new radial coordinate $r=\rho^{1\over 2}$, we obtain
\begin{align}
    \Big( y P(y)\Big)^{1\over 5} \Big({y\over 4 Q(y)} dy^2 +{Q(y)\over P(y)} dz^2\Big)&\sim {y_+^{9\over 5} \over Q'(y_+)} \Big( dr^2 + \left( {Q'(y_+) \over y_+^2}\right) r^2 dz^2\Big)
\end{align}
As discussed above the absence of a conical deficit on the boundary fixes the periodicity of $z$ to be $2\pi$.  
\begin{align}
    {Q'(y_+) \over y_+^2} = {1\over n}
\end{align}
gives us the metric on a half spindle which is regular everywhere except at $y=y_+$ where there is a conical deficit angle $2\pi (1-{1\over n})$. 
\\
Using the explict form of $Q$, we obtain the following constraint on the charges:
\begin{align}\label{nodeficitcondition}
   y_+\Big( 4y_+^2 -(3+{1\over n}) y_+ + 2 (q_1+q_2) \Big) =0
    \end{align}
 Note that the value of the largest root $y_+$ also depends on the charges $q_1,q_2$ and the resulting expression does not have a compact explicit expression. It is however clear that the condition will constrain the charges $q_1,q_2$ to lie on a on dimensional curve, which depends on the value of the conical deficit near the "half-spindle".   In figure \ref{fig:two} we illustrate the curves of allowed charges for the case $n=1$ which corresponds to a completely nonsingular spacetime, and $n=2,3$ which corresponds to 
 spaces with conical deficits $\pi$ and ${2\over 3} \pi$ respectively.
 
We note that there is no completely regular solution with one of the $q_1$ and $q_2$ charges set to zero. Hence all completely regular solutions preserve eight of the thirty two supersymmetries of the $AdS_7$ vacuum of the gauged supergravity. Consequently, the dual four dimensional defect preserves $N=1,d=4$  superconformal symmetry.

\section{Holographic calculations}
\label{sec:four}

The solutions describe holographic co-dimension two defects in the six dimensional SCFT.  In this section we calculate some holographic observables and discuss the implications for the defects imposed by regularity constraints.   As discussed in section \ref{sec:three} the solution approaches $AdS_7$  asymptotically where the six dimensional boundary is $AdS_5\times S^1$. While the boundary is conformal to $R^{1,5}$, it is simpler to work with the $AdS_5\times S^1$  form of the boundary which is natural given the metric (\ref{2dmetric}). All holographic calculations  can be mapped to a flat boundary using the conformal mapping described in section \ref{sec:three}.

\subsection{On shell action}

To evaluate the on shell action we have to add a Gibbons-Hawking term  to the action (\ref{sugract}) which is needed for a good variational principle. Using the trace of the Einstein equation the on-shell action can be expressed as
\begin{align}\label{onshella}
S_{\rm on-shell} &= -{1\over 16 \pi G_N} \int_{M} \sqrt{-g} \Big( {2\over 5} V -{1\over 10}e^{\sqrt{2} \phi_1 + \sqrt{2\over 5} \phi_2 }F_1^2 -{1\over 10}  e^{{-\sqrt{2} \phi_1 + \sqrt{2\over 5} }\phi_2} F_2^2  \Big) \nonumber \\
&\quad  +{1\over 8 \pi G_N} \int_{\partial M} \sqrt{-h} \Theta
\end{align}
The Gibbons-Hawking term is obtained from the trace of the second fundamental form
\begin{align}
\Theta_{\mu\nu}=-{1\over 2}\Big(  \nabla_\mu n_\nu+\nabla_\nu n_\mu\Big)
\end{align}
Here $h_{ab}$ is the induced metric  and $n_\mu$ is the outward pointing normal vector at the  the cut-off surface. For the solution discussed in the paper we choose the cutoff surface at large $y=y_c$. Furthermore since the spacetime closes off at the larges zero $y_+$ of $Q(y)$, the integral  of the coordinate $y$ in the action (\ref{onshella}) is on $y \in [y_+, y_c]$. The on-shell action becomes
\begin{align}
S_{\rm on-shell} &={{\rm Vol}_{AdS_5} \over 16 \pi G_N} \Big( - 10 y_c^3 + 10 y_c^2 -6 (q_1+q_2) y_c -{4\over 5} (q_1+q_2)\nonumber \\
&\quad \quad \quad -{2q_1 q_2\over 5y_+} -{6(q_1+q_2)y_+ \over 5} - 2  y_+^3 + {4\over 5}{ q_1^2\over q_1+ y_+^2}+ {4\over 5}{ q_2^2\over q_2+ y_+^2}\Big)+ o(y_c^{-1})
\end{align}
Here ${\rm Vol}_{AdS_5}$ is the regularized volume of $AdS_5$.
The regularized on shell action is divergent in the limit $y_c\to \infty$ which removes the cutoff. In order to get a finite renormalized action we have to add covariant counter terms at the cutoff surface \cite{Balasubramanian:1999re,Emparan:1999pm,deHaro:2000vlm,Batrachenko:2004fd}
\begin{align}\label{counterterms}
S_{\rm ct}&= {1\over 8 \pi G_N} \int_{y=y_c} \sqrt{-h} \Big(W(\phi_1, \phi_2) +{1\over 8} R[h]+ {1\over 64} (R[h]_{ab}R[h]^{ab}-{3\over 10} R[h]^2)\Big) \nonumber \\
&= {{\rm Vol}_{AdS_5} \over 16 \pi G_N} \Big(  10 y_c^3 - 10 y_c^2 +6 (q_1+q_2) y_c +{5\over 8}\Big)+o(y_c^{-1})
\end{align}
Here $R[h]_{ab},R[h]$ are the Ricci tensor and scalar respectively calculated from the induced metric at the cutoff surface. $W(\phi)$ is the superpotential 
\begin{align}
W(\phi_1,\phi_2)= e^{2 \sqrt{2\over 5}\phi_2}+ 2 e^{-{1\over \sqrt{2}} \phi_1+ {1\over \sqrt{10}} \phi_2}+ 2 e^{-{1\over \sqrt{2}} \phi_1- {1\over \sqrt{10}} \phi_2}
\end{align}
Which is related to the scalar  potential  defined in (\ref{scalpot}) by
\begin{align}
V= 2 \sum_{i=1,2} \left({\partial W\over \partial \phi_i}\right)^2-{6\over 5} W^2
\end{align}
The renormalized action is the given by
\begin{align}
S_{ren}&= \lim_{y_c\to \infty} \Big(S_{\rm on-shell}+ S_{\rm ct}\Big) \nonumber \\
&={{\rm Vol}_{AdS_5} \over 16 \pi G_N} \Big( {5\over 8}-{4\over 5} (q_1+q_2) -{2\over 5}{q_1 q_2\over y_+}- {6\over 5}(q_1+q_2) y_+ - 2  y_+^3+ {4\over 5}{ q_1^2\over q_1+ y_+^2}+ {4\over 5}{ q_2^2\over q_2+ y_+^2}\Big) 
\end{align} 
and when we include the relationship between the $q_i$'s and $y_+$ implied by $Q(y_+)=0$, we obtain a remarkably simple result:
\begin{align}
S_{ren} = {{\rm Vol}_{AdS_5} \over 16 \pi G_N} \big({5\over 8} -2y_+^2 \big)
\end{align}
 
As discussed above, our solutions  describe holographic co-dimension 2 defects. In particular, when $q_1,q_2 = 0$ ($y_+=1$), we just obtain the $AdS_7$ vacuum which must be subtracted in order to identify the quantity above with the expectation value of the defect. 
\begin{align}\label{deltasren}
    S_{ren}-S_{ren}|_{q_1,q_2=0} = {{\rm Vol}_{AdS_5} \over 8 \pi G_N} \big(1- y_+^2\big)
\end{align}
Note that the volume of $AdS_5$ has to be regularized and  will contain a scheme independent logarithmic divergent term. We interpret the coefficient (\ref{deltasren}) as the a central charge \cite{Henningson:1998gx} associated with the four dimensional defect.

\subsection{Stress tensor and currents}
The expectation value of the renormalized holographic stress tensor was derived in \cite{Balasubramanian:1999re,deHaro:2000vlm,Bianchi:2001kw} and can be obtained from the renormalized action
\begin{align}
\langle T_{ab}\rangle_{\rm ren} = {2 \over \sqrt{\rm det(g_{(0)}})} {\partial S_{ren} \over \partial g_{(0)}^{ab} }
\end{align}
Where $g_{(0)}$ is the asymptotic boundary metric in Fefferman-Graham coordinates.
\begin{align}
ds^2&= {d\rho^2\over 4 \rho^2}+ {1\over \rho} g_{ab}(x, \rho) dx^a dx^b
\end{align}
with
\begin{align}\label{fgexp}
 g_{ab}(x, \rho)&= g_{(0),ab} + \rho g_{(2),ab} + \rho^2 g_{(2),ab} + \rho^3 g_{(3),ab} + h_{(3),ab} \rho^3 \log \rho+\cdots
\end{align}
Here the asymptotic boundary is at $\rho=0$.  We defer the details of the calculation to the appendix \ref{appa} 
but note one of the features of the expansion (\ref{fgexp}) is the absence of  the logarithmic term,  i.e. we find $h_{(3),ab}$ vanishes. The final result for the expectation value of the stress tensor is
\begin{align}\label{texpv}
 \langle T_{ab}\rangle_{\rm ren}  dx^a dx^b &= h_D  ds_{AdS_5^2} - 5 h_D ds_{S^1}^2,\quad \quad 
 h_D=\Big( {1\over 18} - {2\over 15}(q_1+q_2) \Big) 
\end{align}
which is traceless, indicating a vanishing six dimensional trace anomaly, which is in accordance with the absence of a logarithmic term in (\ref{fgexp}). The  coefficient $h_D$  can be called the defect's conformal dimension in analogy with other defects such as surface defects in four dimensions \cite{Kapustin:2005py,Jensen:2018rxu,Drukker:2008wr}\footnote{See \cite{Chalabi:2021jud} for an in depth discussion of anomalies for co-dimension two conformal defects.}.

The gauge fields  are dual to conserved currents and from the asymptotic behavior of $A_i$ given in (\ref{2dmetric}),  we can read off the source and expectation value using the standard AdS/CFT dictionary. 
\begin{align}
\lim_{\rho \to 0 } A_i = \Big(q_i \rho^4  +\cdots\Big) dz, \quad i=1,2
\end{align}
which implies that there is no source for the conserved currents and the  expectation value of the currents is given by
\begin{align}
\langle J_i \rangle= q_i dz
\end{align}
Since the currents are dual to the a $U(1)\times U(1)$ R-symmetry, we have a non-vanishing holonomy  around the $S^1$. Recall that the regularity conditions derived in section \ref{sec:three} constrain the charges and hence the holonomies to a one parameter family.

Another holographic observable which can be calculated is the entanglement entropy in the presence of a defect (see e.g.\cite{Jensen:2013lxa,Estes:2014hka,Gentle:2015ruo,Gentle:2015jma}). General arguments relate this quantity to the ones already calculated in this section. \cite{Estes:2018tnu}.

\section{Uplift to 11 dimensions}
\label{sec:five}

The seven dimensional solutions presented in section \ref{sec:two}  can be uplifted to solutions of eleven  dimensional supergravity  \cite{Cvetic:1999xp,Ferrero:2021wvk}
\begin{align} \label{meteleven}
ds_{11}^2&= \Omega^{1\over 3} ds_7^2+ \Omega^{-{2\over 3}} \Big( e^{-\sqrt{8\over5} \phi_2} d\mu_0^2+e^{{\phi_1\over \sqrt{2}}+ {\phi_2\over \sqrt{10}}}(d\mu_1^2+\mu_1^2 (d\phi_1+A_1)^2)\nonumber\\
&\quad  +e^{-{\phi_1\over \sqrt{2}} + {\phi_2\over \sqrt{10}}}(d\mu_2^2+\mu_2^2 (d\phi_2+A_2)^2)\Big)
\end{align}
Where $\Omega$ is defined as
\begin{align}
\Omega= e^{\sqrt{8\over 5}\phi_2} \mu_0^2 + e^{-{\phi_1\over \sqrt{2}}- {\phi_2\over \sqrt{10}}}\mu_1^2+e^{{\phi_1\over \sqrt{2}}- {\phi_2\over \sqrt{10}} }\mu_2^2
\end{align}
The coordinates $\phi_i, i=1,2$ are angular coordinates with periodicity $2\pi$ and the  coordinates $\mu_i$ satisfy the constraint $\sum_{i=0}^2 \mu_i^2=1$. The four form antisymmetric tensor flux is given by
\begin{align}
*_{11} F_4 &= \Big( 2 \sum_{a=0}^2 (X_a^2 \mu_a^2-\Omega X_a) + \Omega X_a\Big) {\rm vol}_7 +{1\over2}\sum_{a=0}^2 {1\over X_a}  (*_7 dX_a) \wedge d(\mu_a^2) \nonumber \\
&\quad +{1\over 2} \sum_{a=1}^2 {1\over X_a^2} d(\mu_a^2)\wedge (d\phi_a+A_a)\wedge *_7 F_a
\end{align}
Here $*_{11}$ is the Hodge dual with respect to the eleven dimensional metric  (\ref{meteleven}) whereas $*_7$ and ${\rm vol}_7$ are the Hodge dual and volume with of  to the seven dimensional metric (\ref {2dmetric}) respectively.   Note that the $AdS_7$ vacuum solution $q_1=q_2=0$ gives the $AdS_7\times S^4$ solution of eleven dimensional supergravity,  dual to the vacuum of the six dimensional SCFT. Since the gauge fields $A_i, i=1,2$ twist the two angular coordinates $\phi_i$ in the metric (\ref{meteleven}) we can identify the gauge fields as dual to $U(1)\times U(1)$ R-symmetry currents inside the $SO(5)$ R-symmetry of the $N=(0,2)$ six dimensional SCFT.

\section{Discussion}

In this note we constructed holographic solutions  of $N=4, d=7$ gauged supergravity which describe four dimensional defects living inside a six-dimensional SCFT.  The solutions are closely related to $AdS_5$ compactifications 
on spindles of the same theory \cite{Ferrero:2021wvk}. The main difference lies in the fact that the two dimensional space  transverse to the $AdS_5$ factor is compact in the spindle case, whereas in our case the space is noncompact and the solution has an asymptotic $AdS_7$ boundary.  For the spindle \cite{Ferrero:2021wvk} the two dimensional space is a sphere with two conical singularities at the north and south pole.  The main  result of the present paper is that for the two charge extremal solutions it is possible to find completely regular solutions without any conical deficits in the bulk or on the asymptotic boundary. These solutions form a one parameter family in the space of extremal solutions.  Another class of solutions are the "half-spindle" solutions of \cite{Bah:2021mzw,Bah:2021hei} where the two dimensional space has the topology of the disk with one conical singularity  in the center and M5-brane sources.  It is possible to generalize our solutions to include a conical singularity in the bulk and in some sense this solution corresponds to a half-spindle on a plane instead of a disk since we have a non-compact space. It would be interesting to investigate whether a relation to the solutions \cite{Bah:2021mzw,Bah:2021hei} exists.  More generally speaking it would be interesting to see whether its possible to modify other holographic solutions of M-theory 
which describe $AdS_5$ compactifications, such as  \cite{Maldacena:2000mw,Gaiotto:2009gz,Gauntlett:2004zh}
  to include a noncompact direction leading to an asymptotic $AdS_7$ boundary and hence describing a defect embedded in a higher dimensional theory.

The asymptotic  boundary of the spacetime is $AdS_5\times S^1$ which is conformal to $R^{1,5}$  under this map the circle parameterizes the angular direction of the transverse $R^2$. Since our solution have a non-vanishing expectation  value of the $U(1)\times U(1)$ R-symmetry currents we can interpret the defect as a homolomy defect for the R-symmetry currents. Examples of such defects have been constructed for free field theories \cite{Aharony:2015zea,Giombi:2021uae,Nishioka:2021uef,Lauria:2020emq,Wang:2021mdq}. For surface defects in four dimensional $N=4$ SYM such defects can be are related  to probe brane and fully back reacted LLM geometries \cite{Lin:2004nb,Lin:2005nh,Gomis:2007fi}
 and some observables were matched in \cite{Drukker:2008wr}. It would be interesting to see whether such a relation exist for four dimensional defects in the six dimensional SCFT, in particular whether there is a field theory analogue of the regularity condition relating the two charges or holonomies that we found.  We leave these interesting questions for future work.

\section*{Acknowledgements}

The work of M. G. was supported, in part, by the National Science Foundation under grant PHY-19-14412. The authors are grateful to the Mani L. Bhaumik Institute for Theoretical Physics for support.

	\newpage
	
	\appendix

	\section{Calculation of holographic stress tensor}
	\label{appa}
In this section we calculate the expectation value of the holographic stress tensor following \cite{deHaro:2000vlm}. The metric (\ref{2dmetric}) has the following large $y$ expansion
\begin{align}\label{dsexp}
 ds^2& = \Big({1 \over  y^2}+ {1\over  y^3} + {5-4(q_1+q_2)\over 5y^4}+\cdots\Big) {dy^2\over 4} + \Big(y+ {q_1+q_2\over 5 y} + {-2 q_1^2-2 q_2^2+ q_1 q_2  \over 25 y^3}+\cdots\Big) ds_{AdS_5}^2 \nonumber\\
&\quad  +\Big( y+ {q_1+q_2\over 5 y} + {4 q_1 +4q_2   \over 5 y^2} + {-2 q_1^2-2 q_2^2+ q_1 q_2  \over 25 y^3}+\cdots\Big) dz^2
\end{align}
	where the dots denote terms which go  faster to zero  in the limit   $y\to \infty$. The following coordinate transformation bring the metric into Fefferman-Graham form
	\begin{align}\label{fgmap}
	y = {1\over \rho} +{1\over 2} + {5- 16(q_1+q_2)\over 80} \rho - { q_1+q_2\over 30} \rho^2 +o(\rho^3)
	\end{align}
	Which takes the following form
	\begin{align}
ds^2&= {d\rho^2\over 4 \rho^2}+ {1\over \rho} g_{ab}(x, \rho) dx^a dx^b  \nonumber \\
 g_{ab}(x, \rho)&= g_{(0),ab} + \rho g_{(2),ab} + \rho^2 g_{(4),ab} + \rho^3 g_{(6),ab} + h_{(6),ab} \rho^3 \log \rho+\cdots
\end{align}
The $g_{ab}$ the takes the following form in Fefferman-Graham coordinates
\begin{align}
g_{ab}(x, \rho)dx^a dx^b &= \Big(1+{1\over 2}\rho+{1\over 16} \rho^2- {2 q_1+2 q_2\over 15}\rho^3 +\cdots\Big) ds_{AdS_5}^2\nonumber\\
&\quad + \Big( 1-{1\over 2} \rho+ {1\over 16} \rho^2+ {2 (q_1+q_2)\over 3} \rho^3+\cdots\Big) dz^2
\end{align}
From which we can read off the $g_{(i),ab}, i=0,2,4,5$. Note that there is no term logarithmic in $\rho$ and hence $ h_{(6),ab}=0$ for the solution considered in this paper. The expectation value of the holographic stress tensor is then given by
\begin{align}
\langle T_{ab}\rangle &=  g_{(6),ab} -A_{(6),ab} + {1\over 24} S_{ab}
\end{align}
Where $A_6$ and $S$ are expressed in terms of $g_{(0)},g_{(2)},g_{(4)}$  and their derivatives. Explict expressions can be found in  \cite{deHaro:2000vlm} and evaluating them for our background gives
\begin{align}
 \langle T_{ab}\rangle_{\rm ren}  dx^a dx^b = \Big( {1\over 18} - {2\over 15}(q_1+q_2) \Big)  ds_{AdS_5^2} +\Big( -{5\over 18} +{2\over 3} (q_1+q_2) \Big) dz^2
\end{align}

	\newpage
\providecommand{\href}[2]{#2}\begingroup\raggedright
\endgroup

\end{document}